\documentclass[a4paper]{jpconf}
\usepackage{graphicx}
\usepackage{hyperref}

\begin{document}
\title{Study of the neutral mesons in Pb-Pb collisions at $\sqrt{s_{NN}}$=2.76~TeV in the ALICE experiment at LHC}

\author{Lucia Leardini on behalf of the ALICE Collaboration}

\address{Physikalisches Institut, Heidelberg University}

\ead{lucia.leardini@cern.ch}

\begin{abstract}
The $\pi^{0}$ and $\eta$ meson production in Pb-Pb collisions at $\sqrt{s_{NN}}$=2.76~TeV is studied with the ALICE experiment at the LHC. The $\pi^{0}$ invariant yields and nuclear modification factor $R\mbox{\tiny{AA}}$ are presented here in six centrality classes. The results are a combined measurement using the Photon Conversion Method (PCM) and the PHOS detector, in the transverse momentum range 0.4 $< p_{\mbox{\tiny{T}}} <$ 12~GeV/$c$. The $\pi^{0}$ $R\mbox{\tiny{AA}}$ is studied in different centrality classes and compared with results from experiments at lower energies, both as a function of transverse momentum. The $\eta$ meson production is studied using the PCM and the EMCal detector. The combination of the individual results will make possible the measurement of the $\eta$ differential invariant cross section as a function of transverse momentum from 1 to 22~GeV/$c$ in different centrality classes.
\end{abstract}

\section{Introduction}
The ALICE experiment \cite{alice} scientific program is focused on the analysis of heavy-ion collisions and on the study of the Quark-Gluon Plasma, a state of the matter with high temperature and large energy density, where gluon and quarks are deconfined, as predicted by QCD. As this medium has high color charge density, partons produced in the earlier stage of the collision lose energy through radiative gluon emission and multiple scattering while traversing it. The parton energy loss affects the hadron yields in the high $p_{\mbox{\tiny{T}}}$ region, a modification that can be observed comparing results on particle production in Pb-Pb collisions with those from pp collisions. In order to study this effect, the nuclear modification factor $R\mbox{\tiny{AA}}$ is used: 
\begin{equation} \label{eq:Raa}
R_{\rm{AA}}(p_\mathrm{T}) = {{\mathrm{d}^2N/dp_\mathrm{T}\mathrm{d}y|_{\rm{AA}}}\over{\langle T_{\rm{AA}}\rangle\mathrm{d}^2\sigma/\mathrm{d}p_\mathrm{T}\mathrm{d}y|_\mathrm{pp}}}
\end{equation} 
where $\langle T_{\rm{AA}}\rangle=\langle N_\mathrm{coll}\rangle/\sigma^\mathrm{pp}_\mathrm{inel}$ is used to scale the pp yield to the Pb-Pb yield. Here, $\langle N_\mathrm{coll}\rangle$ is the average number of binary nucleon-nucleon collisions obtained from the Glauber model \cite{glauber} and $\sigma^\mathrm{pp}_\mathrm{inel}$ is the pp inelastic cross section \cite{ppcrossec}. The $\pi^{0}$ $R\mbox{\tiny{AA}}$ has been studied and is described in the following. The ALICE experiment \cite{perf} measures neutral pions decaying in two photons \cite{ppref}\cite{Abelev}. The detectors involved in the measurement of photons are a lead-scintillator electromagnetic calorimeter (EMCal) and the photon spectrometer (PHOS), made of lead tungstate crystals. The EMCal covers $|\eta| <$~0.7 and $\Delta \phi = 100^\circ$ while PHOS covers $|\eta| <$~0.25 and $\Delta \phi = 60^\circ$. It is also possible to measure photons through their conversion into electrons and positrons in the detector material. This method is called Photon Conversion Method (PCM) and relies on the Inner Tracking System (ITS) and on the Time Projection Chamber (TPC), which have full azimuthal coverage. The material thickness, up to the middle of the TPC, is $X/X_{0}$ = 11.4$\pm$0.5$_{sys}$\%. 
\begin{figure}[h]
\begin{minipage}[b]{18pc}
\includegraphics[width=18pc]{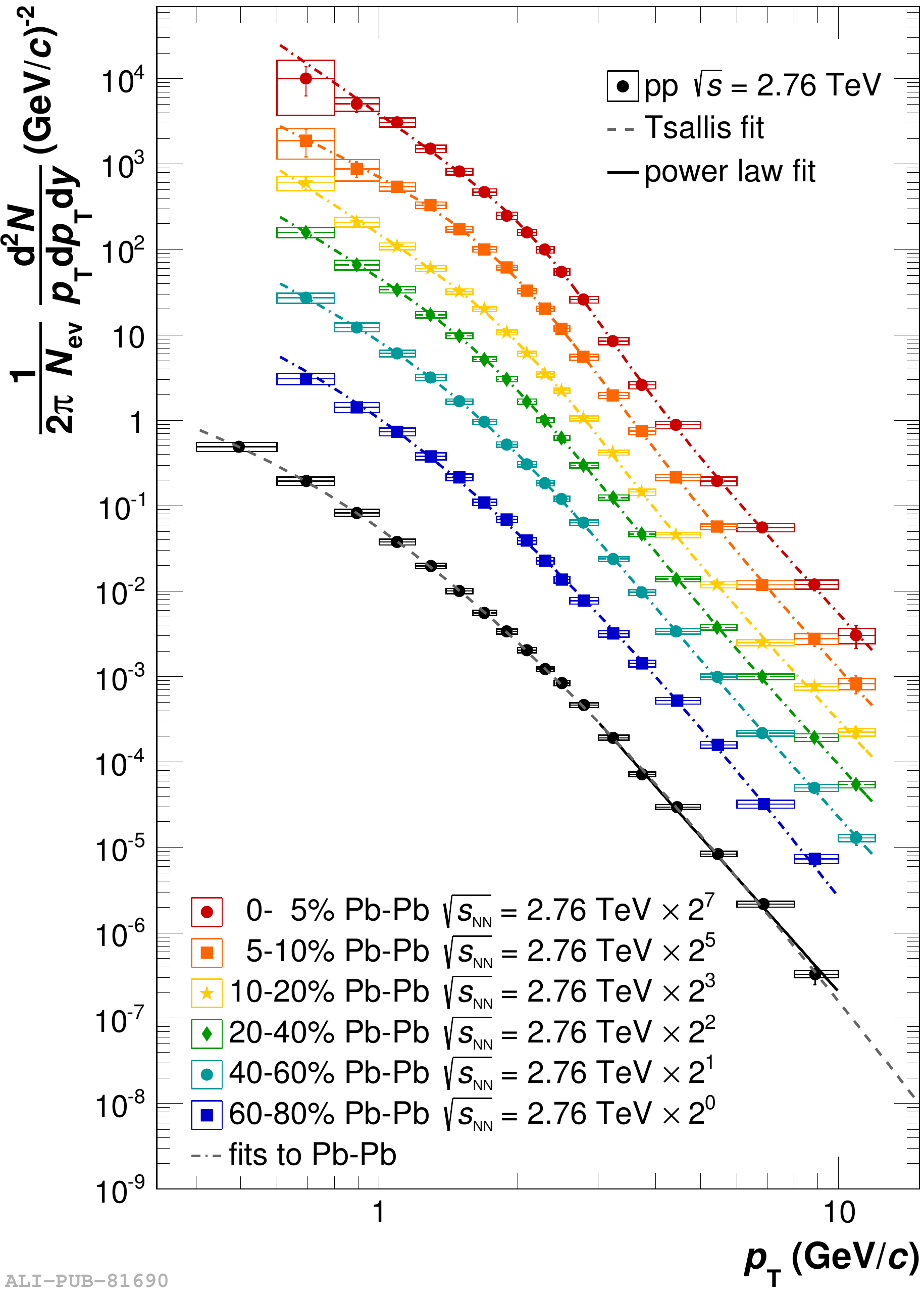}\hspace{2pc}%
\caption{\label{fig:pi0PbPb}(Color online) Invariant differential yields of neutral pions produced in Pb–Pb and inelastic pp collisions. The spectra are the weighted average of PHOS and PCM results. Vertical lines represent statistical uncertainties, boxes systematic uncertainties and horizontal lines indicate bin width.}
\end{minipage}\hspace{2pc}%
\begin{minipage}[b]{18pc}
\includegraphics[width=16pc]{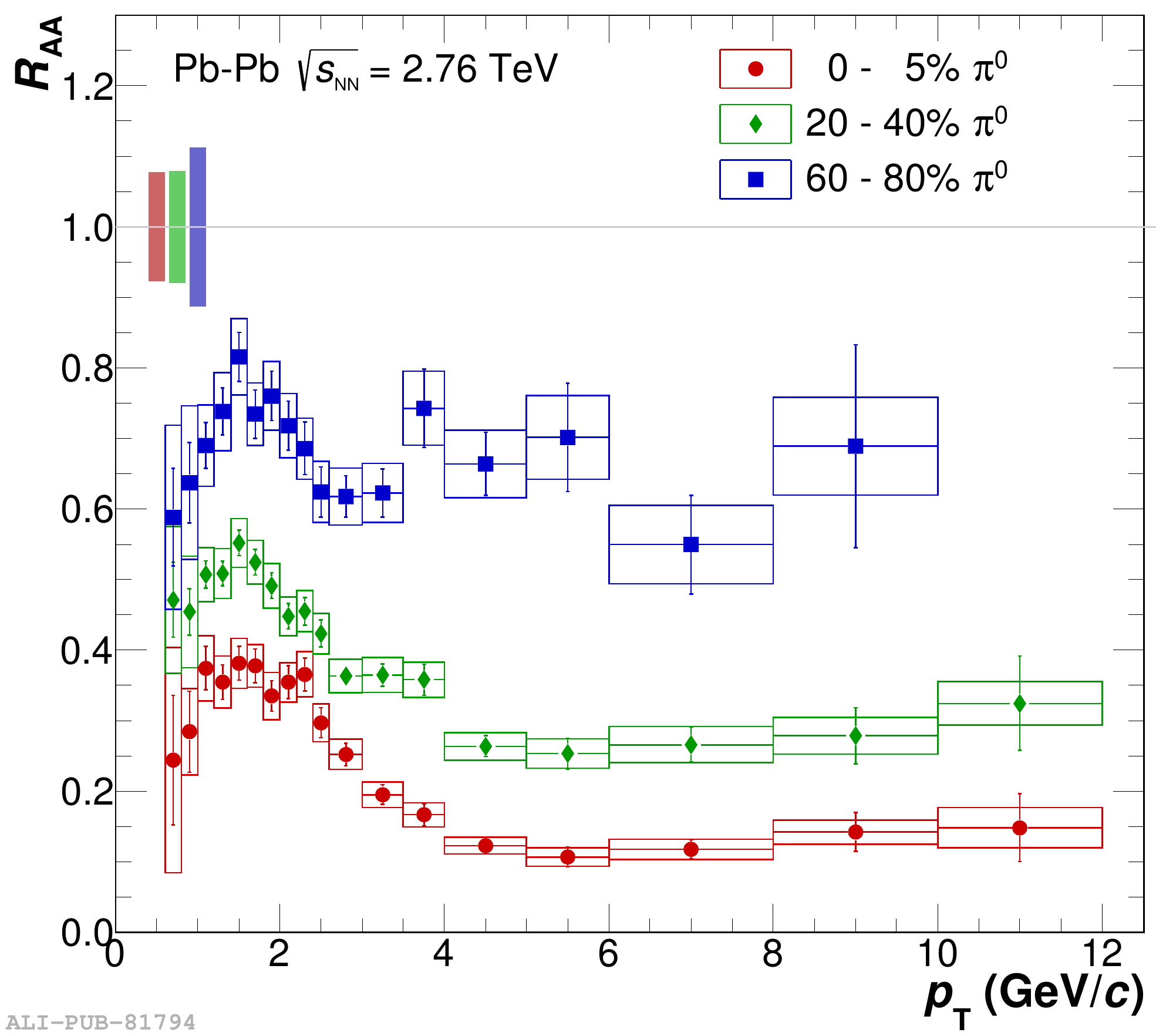}
\vspace{2pc}
\includegraphics[width=16pc]{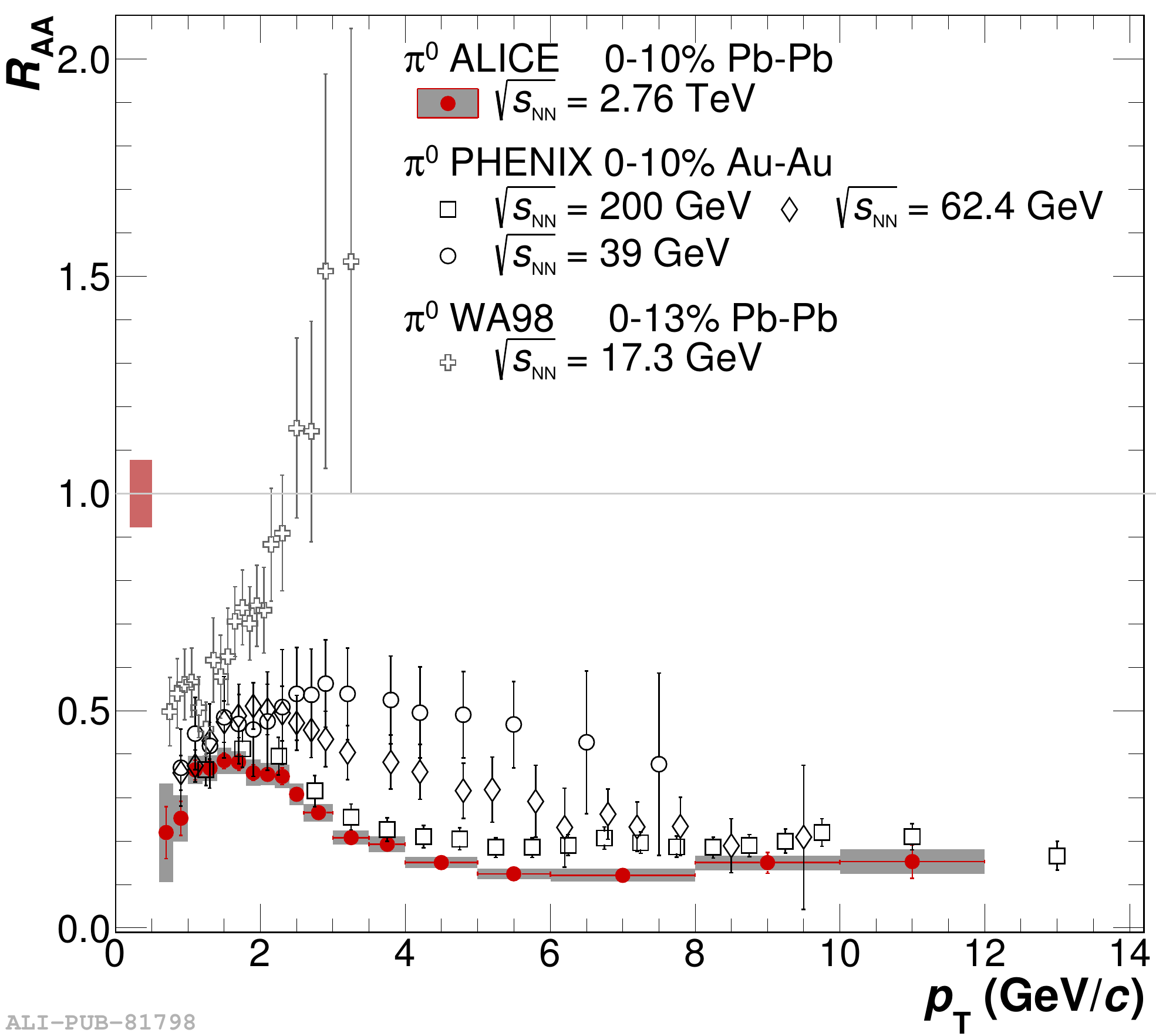}
\vspace{-1pc}
\caption{\label{RAA}(Color online) Top: Neutral pion modification factor in Pb-Pb collisions. Bottom: Neutral pion modification factor in Pb-Pb for centrality class 0-10\% measured by ALICE \cite{Abelev} is compared with the results from Au-Au collisions measured by PHENIX \cite{phenix1}\cite{phenix2} and with the results from Pb-Pb collisions at the CERN SPS \cite{sps}.}
\end{minipage}
\end{figure}

\section{Results}  
The $\pi^0$ invariant yields presented here have been measured independently by PHOS and PCM and were then combined into a weighted average. In Fig.~\ref{fig:pi0PbPb}, the combined PHOS-PCM $\pi^{0}$ yields are shown for pp collisions and Pb-Pb collisions in 6 centrality classes. For PHOS, the data sample consists of 16.1~$\times$~10$^6$ events collected in the 2010 Pb-Pb run and of 34.7~$\times$~10$^6$ events from the 2011 pp run at $\sqrt{s_\mathrm{NN}}=2.76$~TeV, while for PCM it is 13.2~$\times$~10$^6$ and 58~$\times$~10$^6$ events, respectively \cite{Abelev}.
The transverse momentum range goes from 0.6 to 12 GeV/$c$. For the pp spectrum, a Tsallis function \cite{Tsallis} for the full $p_{\mbox{\tiny{T}}}$ region and a power law $E\mbox{d}^{3}N/\mbox{d}^{3}p \propto 1/p^{n}_{\mbox{\tiny{T}}}$, only for the high $p_{\mbox{\tiny{T}}}$ region ($p_{\mbox{\tiny{T}}} >$ 3~GeV/$c$), are shown. The power law fit extrapolation above 8~GeV/$c$ is used as reference in the $R\mbox{\tiny{AA}}$ calculation. The Pb-Pb spectra are fitted with the parametrization described in \cite{Abelev} appendix. \\
The neutral pion nuclear modification factor is calculated with Eq.~\ref{eq:Raa}. The nuclear overlap function $\langle T_{\rm{AA}}\rangle$ was obtained with a Glauber Monte Carlo calculation \cite{glauber}\cite{Glauber1}\cite{Glauber2}, the values and uncertainties for each centrality class can be found in \cite{Abelev}. In Fig.~\ref{RAA}, top panel, the combined PCM-PHOS result for the $\pi^{0}$ $R\mbox{\tiny{AA}}$ is shown. The $R\mbox{\tiny{AA}}$ decreases going from peripheral to central collisions, due to different energy loss in the medium. For all the centrality classes, the $R\mbox{\tiny{AA}}$ has a maximum below 2~GeV/$c$ of transverse momentum and then decreases until, above 6~GeV/$c$, the trend of the data points is flat. In this region, particle production is expected to be dominated by fragmentation of hard-scattered partons. \\
The dependence of $R\mbox{\tiny{AA}}$ on the system energy has also been studied. In the bottom panel of Fig.~\ref{RAA} the $R\mbox{\tiny{AA}}$ for Pb-Pb collisions at $\sqrt{s_\mathrm{NN}}=2.76$~TeV measured with ALICE is compared to previous results at lower energies. ALICE data points are below the lower energies points and their suppression is more evident for $p_{\mbox{\tiny{T}}} >$ 2~GeV/$c$. Models attribute this behavior to the higher initial energy densities created at larger $\sqrt{s_\mathrm{NN}}$. This dominates over the increase of $R\mbox{\tiny{AA}}$ from the harder initial parton $p_{\mbox{\tiny{T}}}$ spectra \cite{zapp} and no increase at high $p_{\mbox{\tiny{T}}}$ is observed. It is also noted that the position of the maximum seems to be shifting towards lower transverse momentum with increasing collision energy. The neutral pion $R\mbox{\tiny{AA}}$ has been compared with different theoretical models. The GLV calculation~\cite{glv1}\cite{glv2} and WHDG~\cite{whdg} prediction describe the interaction of hard-scattered parton with the medium using perturbative QCD. EPOS~\cite{epos} and Nemchik et al.~\cite{nem1}\cite{nem2} calculations use a hydrodynamic description at low $p_{\mbox{\tiny{T}}}$ and absorption of color dipoles at high $p_{\mbox{\tiny{T}}}$. Comparison plots can be found in \cite{Abelev}. \\
\begin{figure}[h]
\centering
\begin{minipage}{14pc}
\includegraphics[width=17.4pc]{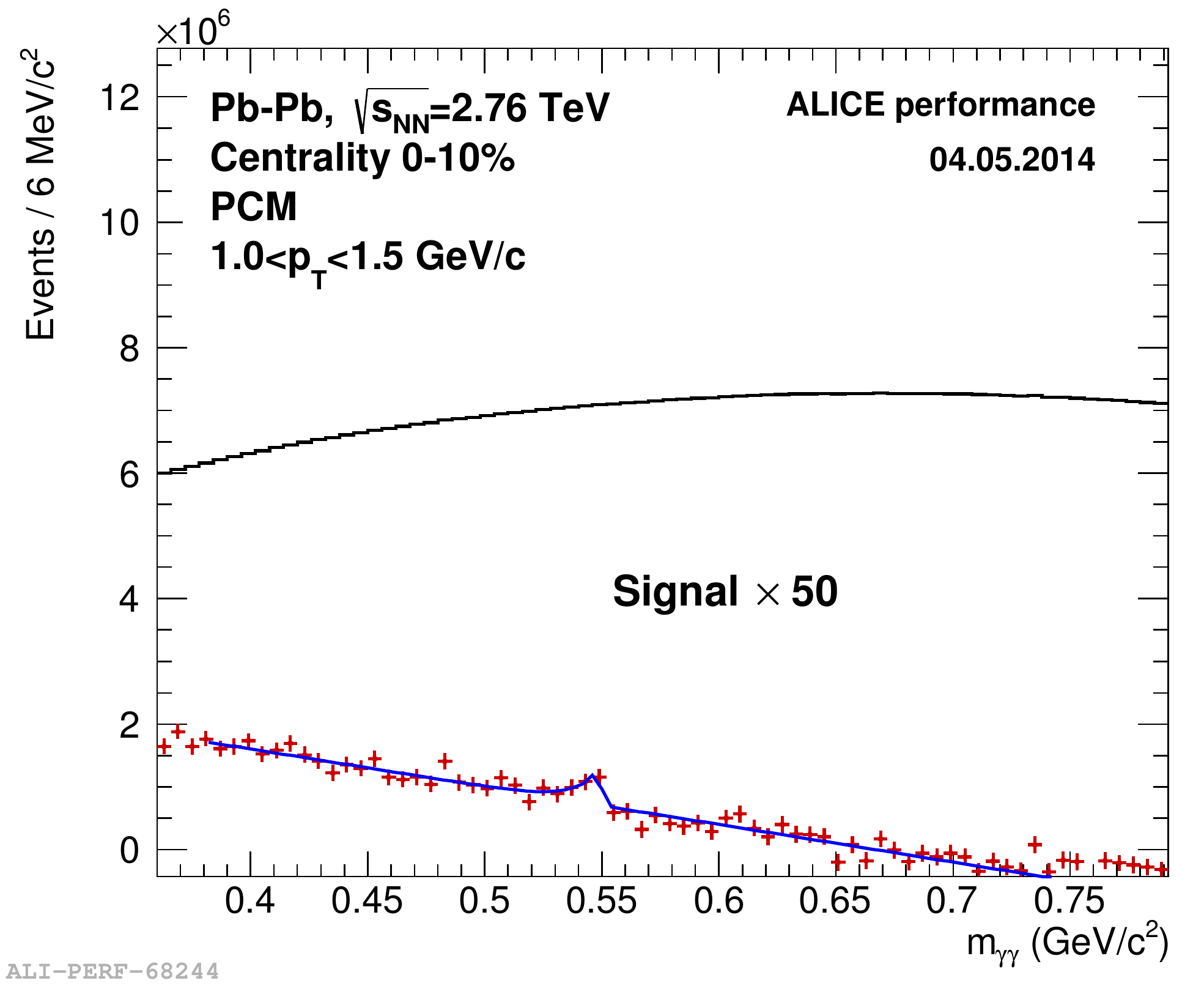}
\end{minipage}\hspace{5pc}%
\begin{minipage}{14pc}
\includegraphics[width=15pc]{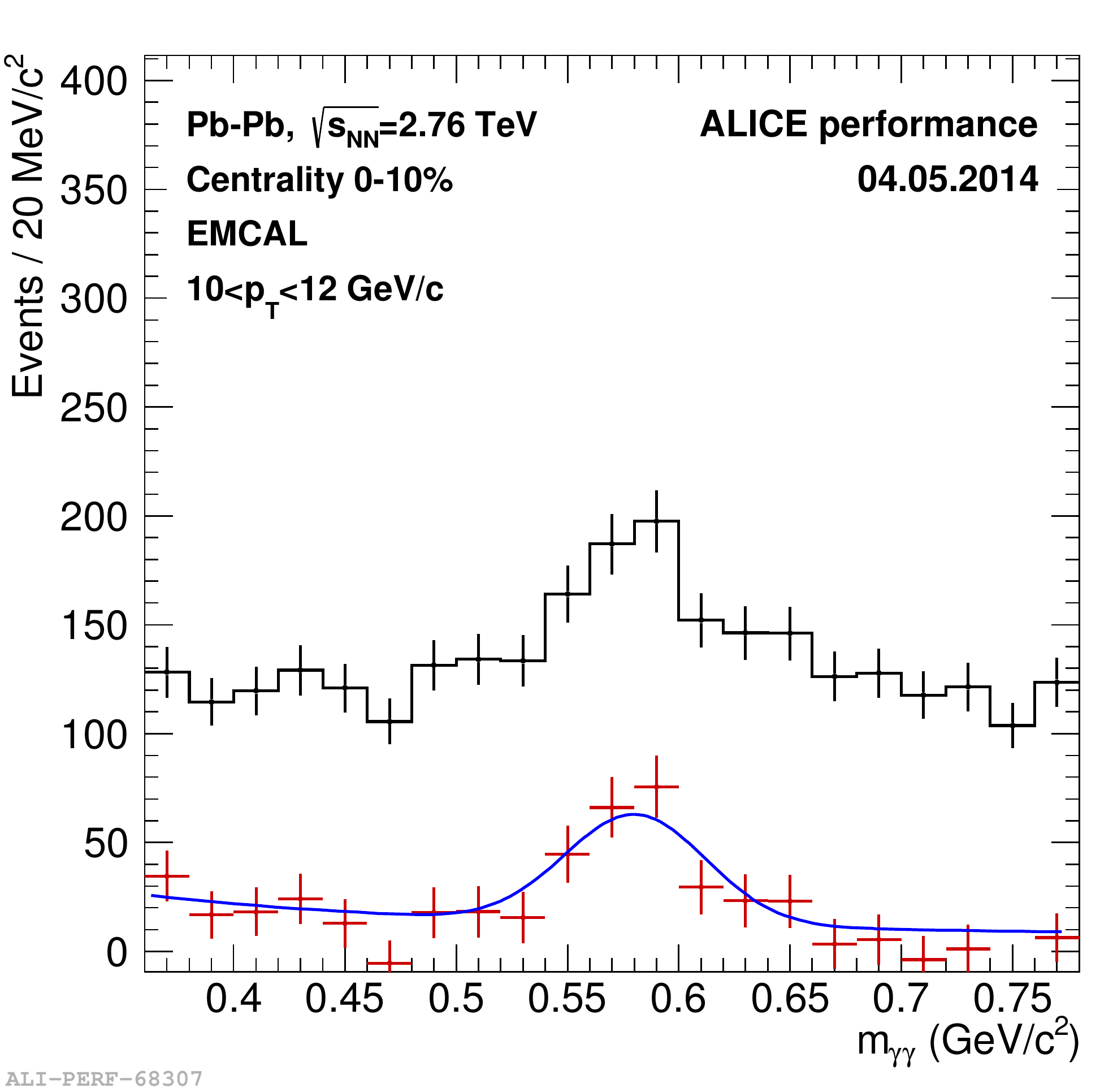}
\end{minipage} 
\caption{\label{eta}Photon pair invariant mass distribution in the $\eta$ mass region in selected $p_{\mbox{\tiny{T}}}$ slices for PCM (left) and EMCal (right) for 0-10\% Pb-Pb collisions. The black histogram and the red points show the data before and after background subtraction, respectively. For PCM, the invariant mass spectra after background subtraction is scaled by a factor 50. The blue line shows the fit to the invariant mass spectra after background subtraction.}
\end{figure}

The $\eta$ meson can be measured in ALICE in Pb-Pb collisions at $\sqrt{s_\mathrm{NN}}=$ 2.76~TeV with the 2011 data sample. It has an integrated luminosity one order of magnitude larger than the corresponding one for 2010, making possible a reliable reconstruction of the $\eta$ meson spectrum. We show here the first steps towards a combined PCM-EMCal $\eta$ meson measurement. The data sample consists of about 17 million events in the central class 0-10\% for both PCM and EMCal. The transverse momentum range for this combined measurement can go from 1 up to 22 GeV/$c$ with an overlap between the two approaches in the intermediate transverse momentum region (4 $<$ $p_{\mbox{\tiny{T}}} <$ 10~GeV/$c$). \\
In Fig.~\ref{eta}, the photon pair invariant mass distribution is shown for two sample transverse momentum bins. The $\eta$ meson peak is clearly visible. For PCM (left panel), the lowest $p_{\mbox{\tiny{T}}}$ bin is shown. The general fit function used by PCM for all $p_{\mbox{\tiny{T}}}$ is a Gaussian function combined with an exponential tail on left side of the peak (to account for the electron bremsstrahlung) and with a linear fit to describe the remaining combinatorial background under the peak after the combinatorial background subtraction. The fit is used only to extract the $\eta$ peak position and width. The $\eta$ yields are then extracted by integrating the peak area and subtracting the integral of the linear fit.
In the right panel of Fig.~\ref{eta}, a bin in the intermediate range of the EMCal $p_{\mbox{\tiny{T}}}$ reach is shown. The fit function used here is a Crystal-Ball function while a linear fit or a parabola, depending on the $p_{\mbox{\tiny{T}}}$ and centrality, is adopted for the residual background hypothesis.

\section{Conclusions}
Measurements of the neutral pion production in Pb-Pb collisions at $\sqrt{s_\mathrm{NN}}=$ 2.76~TeV have been shown combining two independent methods, the measurement of the photons with the PHOS detector and the reconstruction of the converted photons with the tracking system. These two independent measurements are consistent and were combined in a weighted average for the $\pi^{0}$ spectra. The neutral pion nuclear suppression factor $R\mbox{\tiny{AA}}$ was calculated from the measured neutral pion spectra separately for the two methods and then combined, in order to reduce the systematic errors. The $\pi^{0}$ $R\mbox{\tiny{AA}}$ was studied for different centrality classes and compared to low-energy results. The suppression is stronger for higher collision energy when comparing Pb-Pb collisions at $\sqrt{s_\mathrm{NN}}=$ 2.76~TeV to results from RHIC and CERN SPS. \\
The $\eta$ meson invariant mass distribution for two transverse momentum bins from the 2011 Pb-Pb collisions at $\sqrt{s_\mathrm{NN}}=$ 2.76~TeV is shown. Also here two independent methods are used, photon measurement with the EMCal detector and photon conversion. The transverse momentum range for this combined measurement will go from 1 up to 22 GeV/$c$ with an overlap in the intermediate transverse momentum region (4 $<$ $p_{\mbox{\tiny{T}}} <$ 10~GeV/$c$). This analysis will allow for a study of the $\eta$ meson suppression in Pb-Pb collisions.

\end{document}